\documentclass{PoS}
\usepackage{color,graphicx}
\usepackage{subfigure}
\usepackage{amsmath}

\graphicspath{{../../QNP2012/proceedings/}}

\title{A Partial-Wave Analysis of Centrally Produced Two-Pseudoscalar Final States in {\em pp} Reactions at COMPASS}

\ShortTitle{A Partial-Wave Analysis of Centrally Produced Two-Pseudoscalar Final States in $pp$ Reactions at COMPASS}

\author{\speaker{Alexander Austregesilo}%
  \thanks{The author acknowledges financial support by the German Bundesministerium f\"ur Bildung und Forschung (BMBF), by the Maier-Leibnitz-Laboratorium der LMU und TU M\"unchen, and by the DFG cluster of excellence 'Origin and Structure of the Universe'.}\\
        Technische Universit\"at M\"unchen\\
        E-mail: \email{aaust@cern.ch}}

\author{for the COMPASS collaboration}

\abstract{COMPASS is a fixed-target experiment at CERN SPS which 
focused on light-quark hadron spectroscopy during the data taking in 2008 and 2009. A world-leading data set was collected with a 190\,GeV/$c$ hadron beam impinging on a liquid hydrogen target in order to study the central production of glueball candidates.

In this report, we motivate double-Pomeron exchange as a relevant production process for mesons without valence quark content. We select a centrally produced sample from the COMPASS data set recorded with a proton beam and introduce a decomposition into partial waves. Particular attention is paid to inherent mathematical ambiguities in the amplitude analysis of two-pseudoscalar final states. Furthermore, we show a simple parametrisation for the centrally produced $K^+K^-$ system which can describe the mass dependence of the fit results with sensible Breit-Wigner parameters.}

\FullConference{51st International Winter Meeting on Nuclear Physics\\
   21-25 January 2013\\
    Bormio (Italy)}

\begin{document}

  \section{Introduction}
  \label{sec:mot}

\begin{figure}[ht]
    \begin{minipage}{.55\textwidth}
      \begin{center}
        \vspace{.5cm}
        \includegraphics[width=\textwidth]{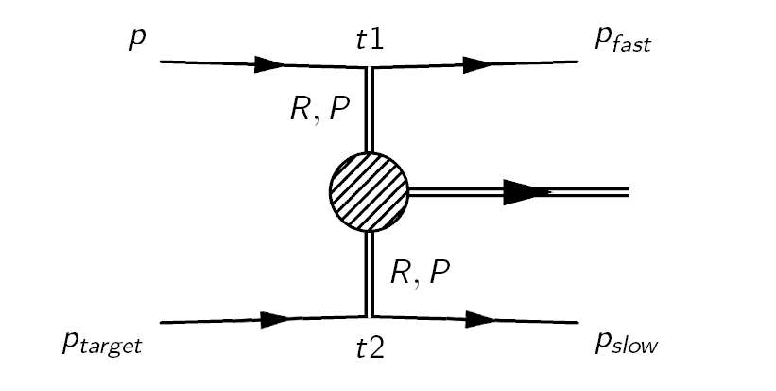}
        \vspace{-.5cm}
        \caption{\em  Central production.}
        \label{fig:cp}
      \end{center}
    \end{minipage}
    \begin{minipage}{.44\textwidth}
      \begin{center}
        \includegraphics[width=\textwidth]{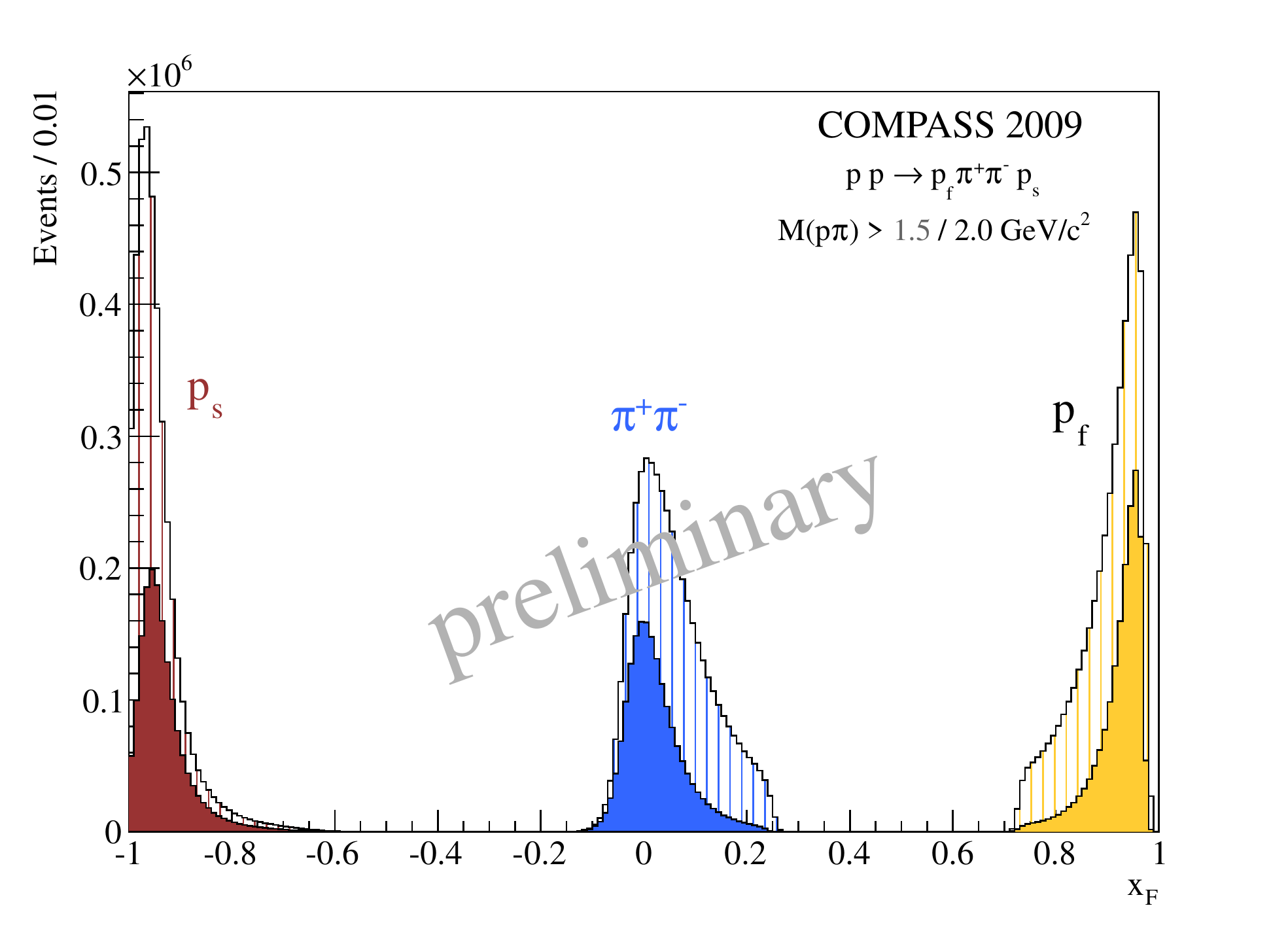}
        \caption[{\em Feynman $x_F$}]{\em  Feynman $x_F$ distributions.}
        \label{fig:xf}
      \end{center}
    \end{minipage}
  \end{figure}

  Quantum Chromodynamics predicts objects composed entirely of valence gluons, so-called {\em glueballs}. Their existence, however, could not be confirmed experimentally to the present day. One of the goals of the COMPASS experiment~\cite{com07} at CERN is to study the existence and signatures of glueballs, in continuation of the efforts that were made at the CERN Omega spectrometer in the late 1990s~\cite{kir97}. Since Pomerons are considered to have no valence quark contribution, Pomeron-Pomeron fusion was proposed to be well suited for the production of glueballs. This process can be realised in a fixed-target experiment by the scattering of a proton beam on a proton target, where a system of particles is produced at central rapidities (cf. Figure~\ref{fig:cp}).

A $190\,\textrm{GeV}/c$ proton beam impinging on a liquid hydrogen target was used for the presented analysis. The trigger on the recoil proton resulted in a large rapidity gap between the slow proton $p_s$ and the other final-state particles measured in the forward spectrometer. Additional kinematic cuts where used in order to separate the central two-pseudoscalar system from the fast proton $p_f$. For the di-pion system for example, a cut on the invariant mass combinations $M(p\pi) > 1.5\,\mathrm{GeV}/c^2$ was introduced. The result of this selection on the Feynman $x_F$ distributions for the fast (yellow) and the slow proton (red) as well as the di-pion system (blue) is shown in Figure~\ref{fig:xf}. The $\pi^+\pi^-$ system  lies within $\left \vert x_F \right \vert \le 0.25$ and can therefore be considered as centrally produced.

\begin{figure}[ht]
    \begin{center}
      \subfigure[{\em $\sqrt s = 12.7\,\mathrm{GeV}/c^2$}~\cite{arm91}]{\includegraphics[width=.25\textwidth]{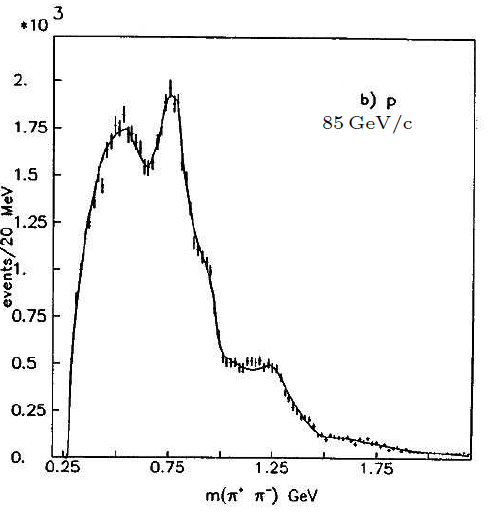}}
      \subfigure[{\em $\sqrt s = 18.9\,\mathrm{GeV}/c^2$}]{\includegraphics[width=.4\textwidth]{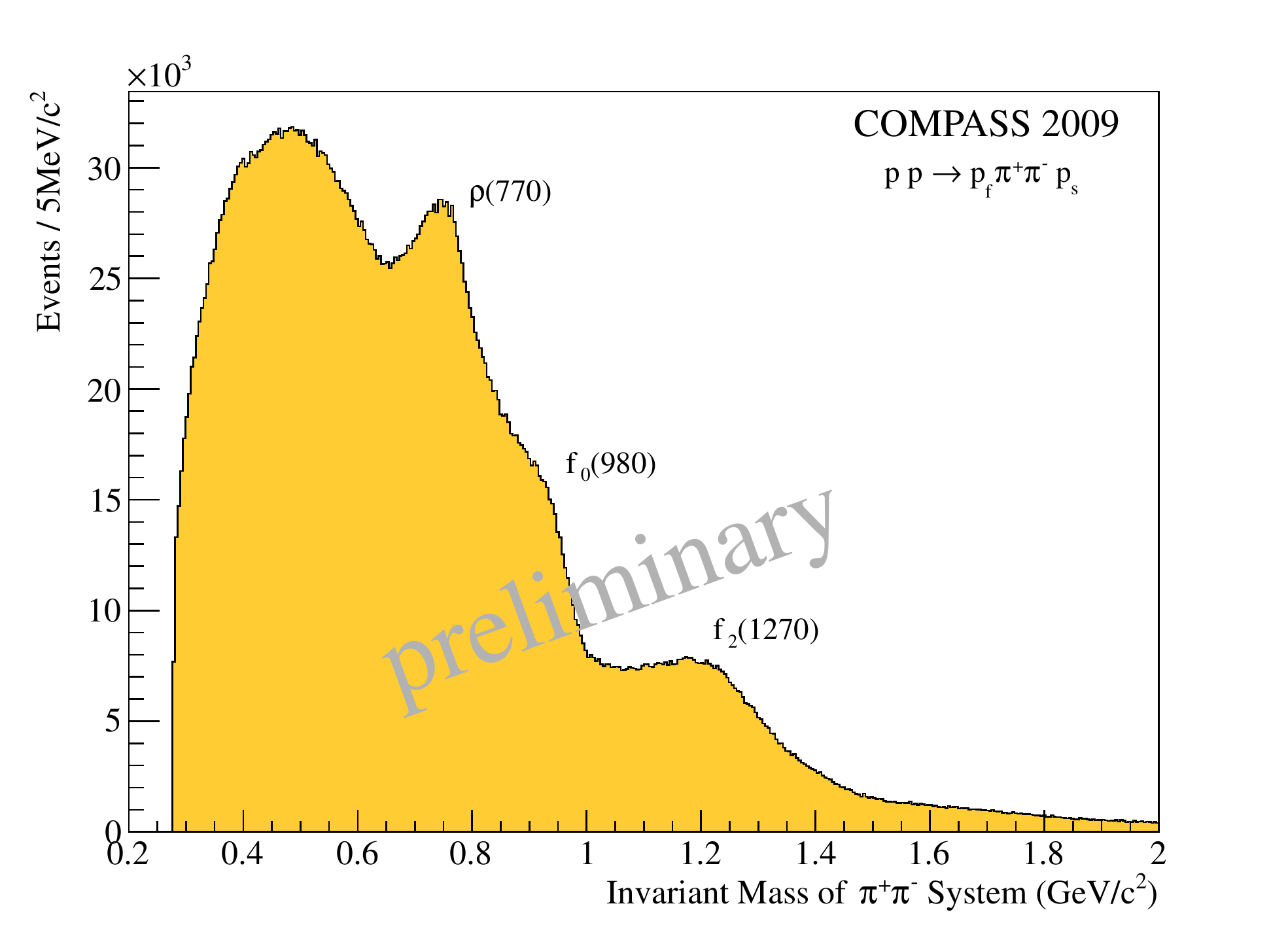}}
      \subfigure[{\em $\sqrt s = 23.7\,\mathrm{GeV}/c^2$}~\cite{arm91}]{\includegraphics[width=.265\textwidth]{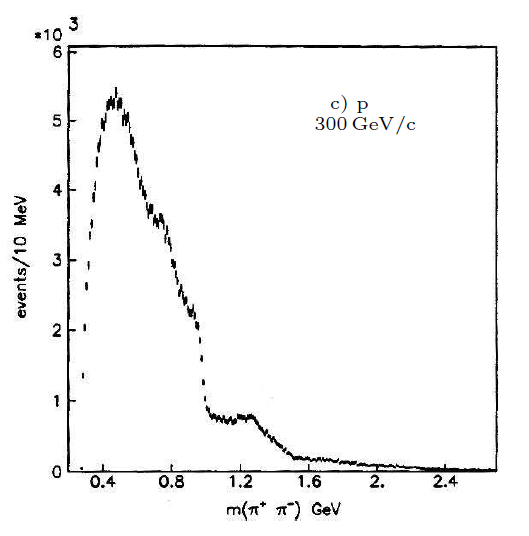}}
    \end{center}
    \caption[{\em $\pi^+\pi^-$ Invariant Mass}]{\em Invariant mass spectrum of the $\pi^+\pi^-$ system measured with Omega (a,c) and COMPASS (b).}
    \label{fig:two}
  \end{figure}

Central production comprises several production processes, their characteristic signature being the different dependence of the cross section on the centre-of-mass energy $\sqrt s$ of the reaction. While Pomeron-Pomeron scattering should be only weakly dependent on $s$, Pomeron-Reggeon scattering is predicted to scale with $1/\sqrt s$ and Reggeon-Reggeon scattering with $1/s$~\cite{kle07}. This behaviour can be observed experimentally, e.g. by comparing the mass spectra obtained at COMPASS energy to those the Omega spectrometer~\cite{arm91} measured at different energies (cf. Figure~\ref{fig:two}).
The dominant features of the $\pi^+\pi^-$ invariant mass distribution, the $\rho$(770), the $f_2$(1270), and the sharp drop in intensity in the vicinity of the $f_0$(980), can be observed at all three different centre-of-mass energies. The relative yield of $\rho$(770)-production decreased rapidly with increasing $\sqrt s$ since it cannot be produced via double-Pomeron exchange. On the other hand, 
the enhancement at low masses as well as the $f_0$(980) remain practically unchanged; a fact that is characteristic for $s$-independent Pomeron-Pomeron scattering.

  \section{Partial-Wave Analysis}

The partial-wave analysis has been performed assuming that the central two-pseudoscalar system is produced by the collision of two particles emitted by the scattered protons. These exchange particles carry the squared four-momentum transfer $t_1$ from the beam proton and $t_2$ from the target proton, respectively.

\begin{figure}[ht]
    \subfigure[{\em $\cos\theta$}]{\includegraphics[width=.45\textwidth]{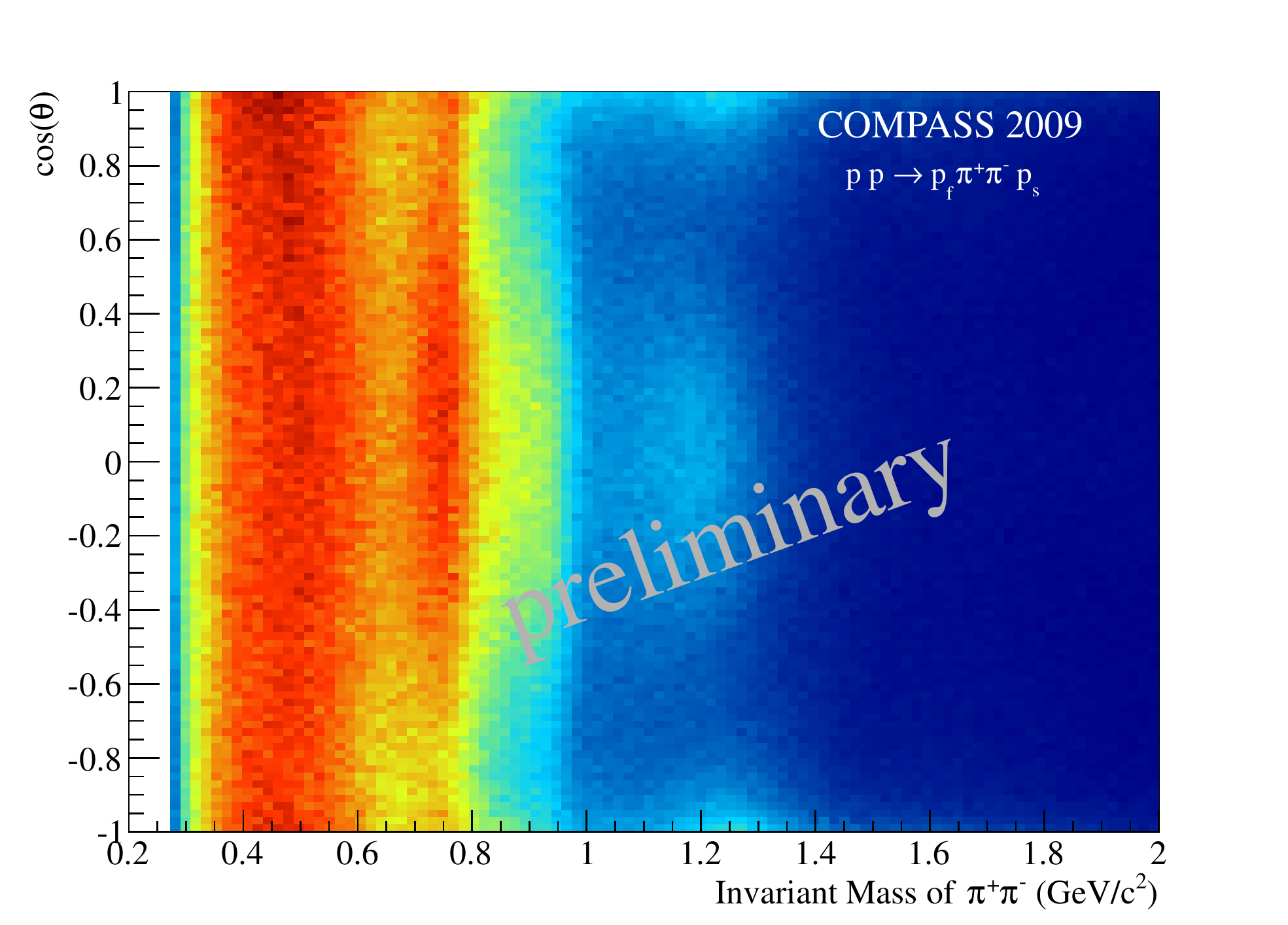}}
    \subfigure[{\em $\phi$}]{\includegraphics[width=.45\textwidth]{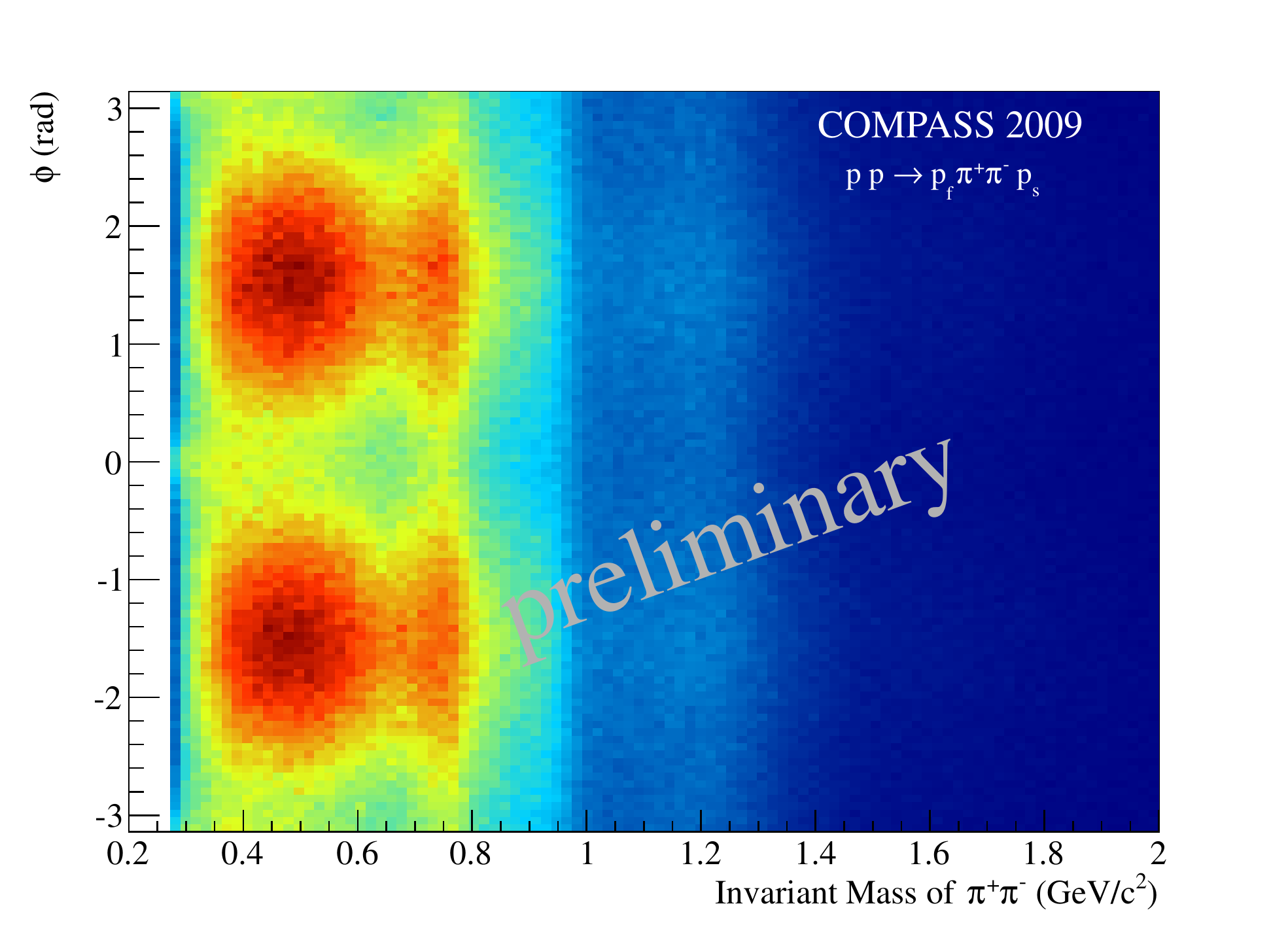}}
    \caption{\em  Decay angles as a function of the $\pi^+\pi^-$ mass.}
    \label{fig:pwa}
  \end{figure}

\begin{floatingfigure}[l]
  \includegraphics[width=4cm]{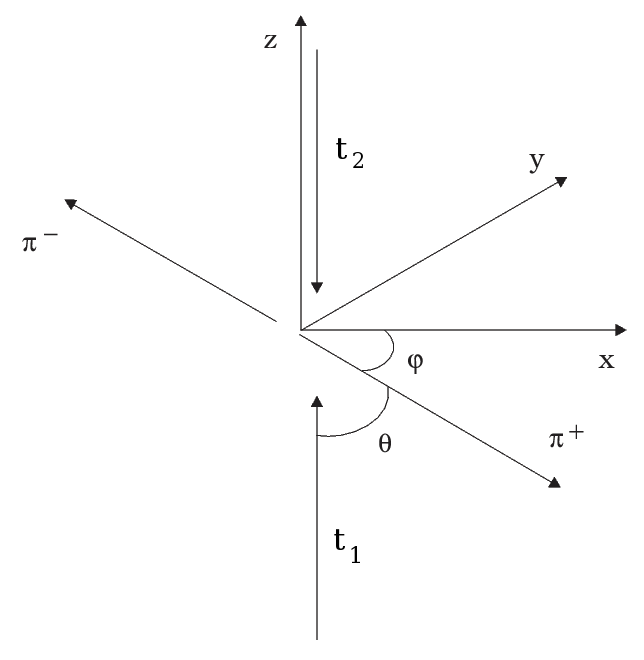}
  \caption{Coordinate system in $\pi^+\pi^-$ centre-of-mass}
  \label{}
\end{floatingfigure}
We make the strong assumption that $t_1$ only transmits the helicity $\lambda = 0$. In this limit, $t_1$ can be seen as a vacuum-like Pomeron with parity $P=+1$, which is treated like an {\em external particle}. The beam and the fast outgoing proton $p_f$ are ignored in this picture. If we accept the space-like Pomeron as the incoming beam, we can construct a Gottfried-Jackson frame~\cite{got64} for the centrally produced system. Hence, the exchanged particle $t_1$ in the centre-of-mass frame of the $\pi^+\pi^-$ system defines the $z$-axis for the reaction. We have to emphasise that we fixed the choice to $t_1$ in contrast to previous experiments (e.g. \cite{bar99}) to be able to correct for the different $t$-acceptances of the trigger. The $y$-axis of the right-handed coordinate system is defined by the cross product of the momentum vectors of the two exchange particles in the $pp$ centre-of-mass system. Apart from the invariant mass, two additional variables specify the decay process; namely the polar and azimuthal angles $\cos\theta$ and  $\phi$ of the negative particle in the two-pseudoscalar centre-of-mass frame relative to these axes.
  Figure~\ref{fig:pwa} illustrates the distribution of these decay variables as a function of the $\pi^+\pi^-$ mass.

  The decay amplitudes with relative orbital angular momentum $\ell$ and its projection on the quantisation axis $m$ is given by the spherical harmonics $Y^\ell_m(\theta, \phi)$. To profit from the fact that the strong interaction conserves parity, a reflection operator was introduced in \cite{chu75} as a parity operator followed by a rotation around the $y$-axis such that all momenta relevant to the production process are recovered. Introducing a quantum number $\varepsilon$ which is called {\em reflectivity} and which can have the values $\pm 1$, the eigenstates of this reflection operator can be constructed as:
  \begin{equation}
    Y^{\ell \varepsilon}_m(\theta, \phi) \equiv c_m \left [ Y^\ell_m (\theta , \phi) - \varepsilon (-1)^m Y^\ell_{-m}(\theta, \phi) \right ]
    \label{eq:eigen}
  \end{equation}
  with a normalisation constant $c_m$. The two classes of states $\varepsilon = \pm1$ correspond to different production processes in the asymptotic limit of high energy and low momentum transfer~\cite{got64} and can consequently not interfere.

 With the complex {\em transition amplitudes} $T^{\varepsilon}_{\ell m}$ and an explicit incoherent sum over the reflectivities, the intensity in narrow mass bins can be expanded in terms of partial waves:
  \begin{equation}
    I(\theta, \phi) = \sum_\varepsilon \left \vert \sum_{\ell m} T^{\varepsilon}_{\ell m} Y^{\ell \varepsilon}_m(\theta, \phi) \right \vert^2
  \end{equation}
  
  We adopt the spectroscopic notation $\ell^\varepsilon_m$ from \cite{bar99} to construct the basic wave set $\mathbf{S^-_0}$, $P^-_0$, $P^-_1$, $D^-_0$, $D^-_1$ and $\mathbf{P^+_1}, D^+_1$. Since the overall phase for each reflectivity is indeterminate, one wave in each class can be defined real, which means the imaginary part of that transition amplitude is fixed to zero. These so-called {\em anchor} waves are marked in bold font.

  An {\em extended maximum-likelihood} fit in $10\,\mathrm{MeV}/c^{2}$ mass bins is used to find the parameters $T_{\varepsilon l m}$, such that the acceptance corrected model $I(\theta, \phi)$ matches the measured data best. We want to stress, that the transition amplitudes are assumed to be constant over the narrow mass bins. In other words, we do not assume any mass dependence in this model-independent decomposition. Only as a second step~(cf.~Section~\ref{sec:massdep}), we will try to interpret the results of this fit with a parametrisation in terms of Breit-Wigner functions.


  \section{Ambiguities}

As shown in \cite{sad91,chu97}, the dependency of the intensity $I(\Omega)$ for two-pseudoscalar final states on the polar angle $\theta$ can generally be expressed in terms of $|G(u)|^2$ by the introduction of a variable $u \equiv \tan(\theta/2)$ ({\em Weierstrass substitution}). The function $G(u)$ is a complex polynomial of the order of $2\ell$, where $\ell$ is the the highest considered spin in the system. This polynomial can be factorised in terms of its complex roots  $u_{k}$ ($k \in [0,..,2\ell]$), the so-called {\em Barrelet-zeros}~\cite{bar72}. Since the function $G(u)$ only enters as absolute square in the expression for the angular distribution, the complex conjugate of a root $u_k^*$ is an equally valid solution. That means, that there are in general $2^{2\ell-1}$ different mathematically ambiguous solutions which result in exactly the same angular distribution. This ambiguity has to be resolved by physical arguments.

\begin{figure}[t]
    \begin{center}
      \subfigure[]{
        \includegraphics[width=.45\textwidth]{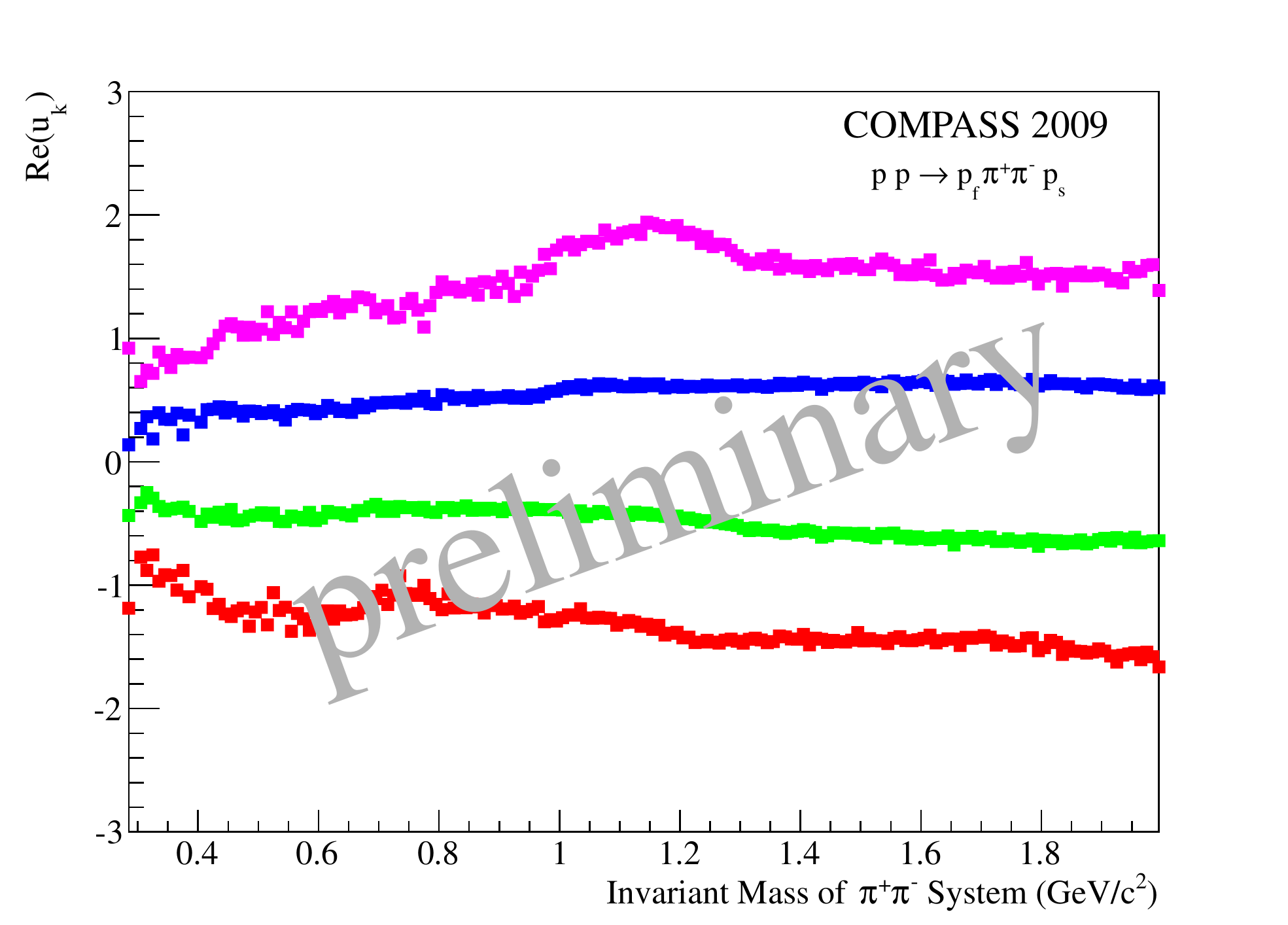}
      }
      \subfigure[]{
        \includegraphics[width=.45\textwidth]{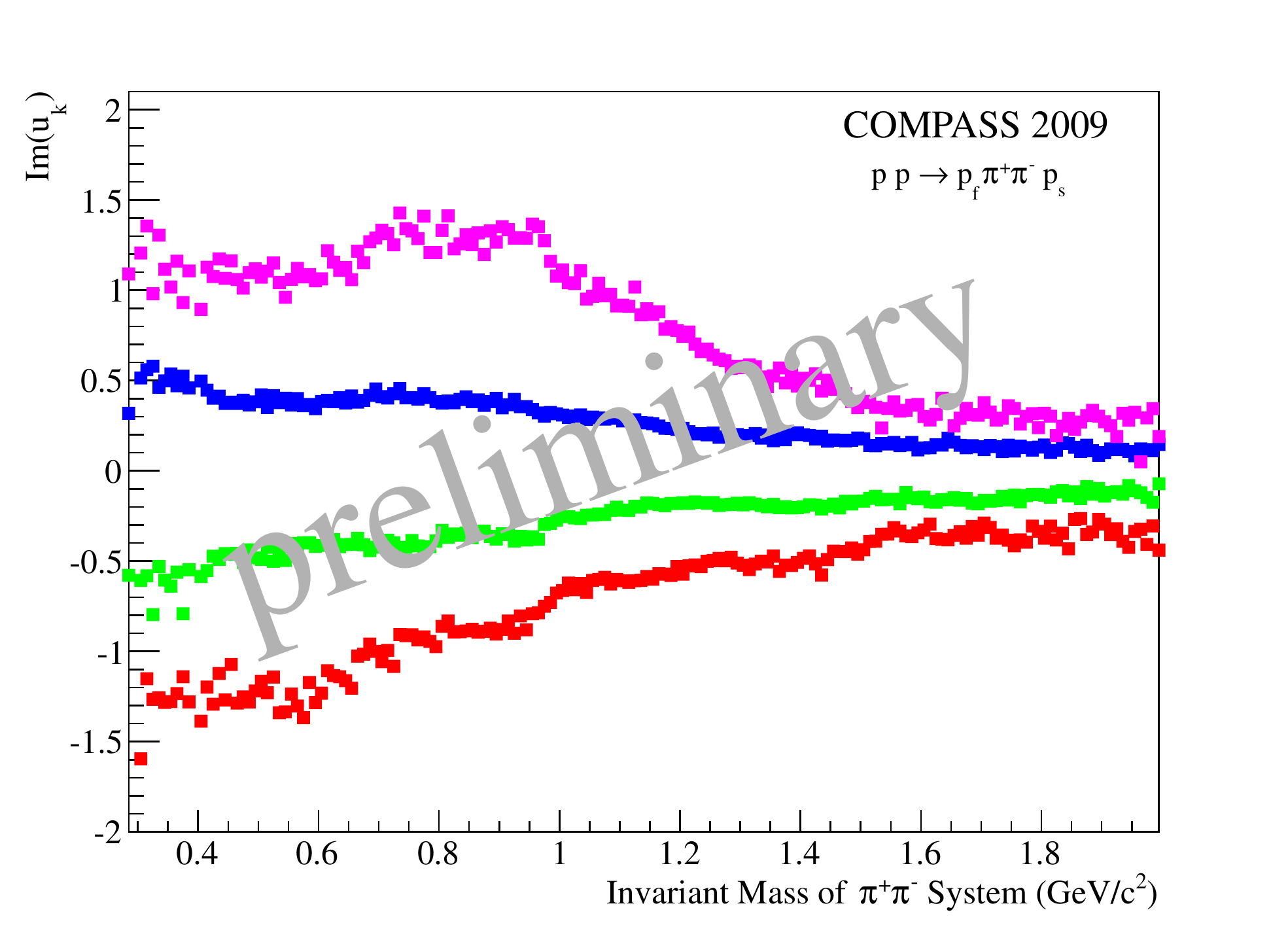}
      }
    \end{center}
    \caption[{\em Barrelet Zeros}]{{\em  The (a) real and (b) imaginary parts of the Barrelet-zeros as a function of the $\pi^+\pi^-$ mass.}}
    \label{fig:roots}
  \end{figure}

The system of $S$, $P$ and $D$ waves used in the presented analysis has eight ambiguous solutions. By using different random starting values for the fit to the same data and MC sample, it was shown experimentally that very different solutions can be obtained. However, the fitted production amplitudes $T_{\varepsilon l m}$ for one single attempt can be used to calculate all eight solutions analytically through the complex polynomial roots~\cite{chu97}. {\em Laguerre's method} was applied in order to find these numerically. Figure~\ref{fig:roots} illustrates the real and imaginary parts of the four roots for all mass bins of the $\pi^+\pi^-$-system, where a sorting depending on the real part has been performed. They are well separated from each other and can be easily linked from mass bin to mass bin. The imaginary parts do not cross the real axis, hence bifurcation of the solutions does not pose a problem and the solutions can be uniquely identified. 

  By fixing $u_1$ and allowing $u_{2,3,4}$ to undergo complex conjugation, the entire set of eight ambiguous solutions is calculated. In order to avoid involved error propagation, these calculated values for one solution are reintroduced to the fit as starting values, which probes the convergence and provides the correct covariance matrix.

 Differentiation among these mathematically equivalent solutions requires additional input, e.g. the behaviour at threshold or the expected physical content. However, the choice is not evident for the $\pi^+\pi^-$ system. Four solutions can be clearly ruled out, since most of the intensity is formed by one single wave. The others remain subject to further studies, where different final states and the mass-dependent parametrisation may bring extra input.

  \section{Evaluation of the Fit with Weighted Monte-Carlo Sample}
  \label{sec:weighted}
  
  To evaluate the fit quality, the decay amplitudes $Y^{\ell \varepsilon}_m$ of a phase-space Monte-Carlo sample are weighted by the production amplitudes $T_i$ obtained in the minimisation:
  \begin{equation}
    \sigma_i = \frac{4\pi}{N_{\mathrm{MC}}} \sum_{\varepsilon} \left \vert \sum_{\ell m} T^{\varepsilon}_{\ell m} Y^{\ell \varepsilon}_m(\theta_i, \phi_i) \right \vert ^2
  \end{equation}
  The acceptance of the apparatus is taken into account by setting the weight of events that did not pass the selection to zero. The factor $4\pi$ cancels the solid angle integral, since the $Y^{\ell \varepsilon}_m$ are normalised to unity. The definition of the production amplitudes for the extended log-likelihood fit includes a normalisation to the number of events in the data. Therefore, respecting the numbers of events in the MC sample, the resulting distributions can be directly compared to the measured ones.
  
\begin{figure}[ht]
    \begin{center}
      \includegraphics[width=.8\textwidth]{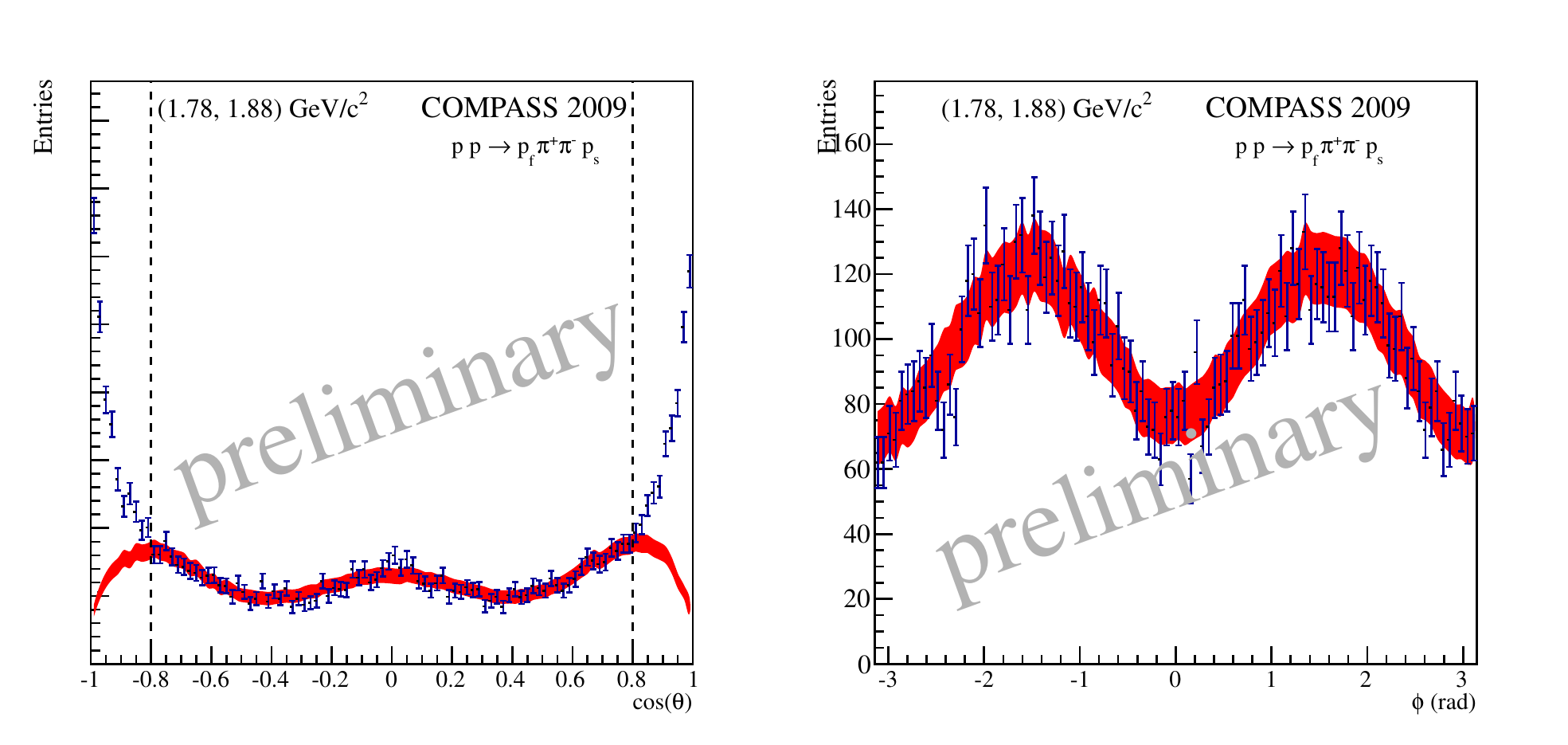}
    \end{center}
    \caption[{\em {\bf Release:} Data to Weighted MC Comparison for Four Mass Ranges}]{\em  Real data (blue) and weighted Monte-Carlo (red) for a di-pion mass of $[1.78,1.88]\,\mathrm{GeV}/c^2$. \\ The cut on $|\cos\theta| < 0.8$ (see text) is indicated with a dashed line.}
    \label{fig:predict}
  \end{figure}

  In general, the $\cos\theta$ and $\phi$ distributions of the fit model show very good agreement with the data. However, for masses above $1.5\,\mathrm{GeV}/c^2$, shortcomings of the model become apparent. A strong forward-backward peaking of the data can be observed (cf.~Figure~\ref{fig:predict}) which cannot be reproduced by the limited wave set. 
  This behaviour can be explained by background from different production processes. For example, a fast baryon resonance, producing a central pion via Regge-exchange as expected by~\cite{kle07}, would have exactly this signature. 
  The fraction of baryon resonances to protons directly depends on the fraction of Regge-exchange, an interesting transition region COMPASS is able to explore quantitatively.
  
  Since the simple model of pure central production could not describe the data in this kinematic region, we decided to perform the whole analysis with a cut on $|\cos\theta| < 0.8$. All presented results were produced with this restriction, the data reduction amounted to $\approx 20\%$. It was verified that the cut does not influence the  results qualitatively. However, the comparison of weighted Monte-Carlo and data shows significant improvements in the higher mass regions.
  
  \section{Partial-Wave Analysis of the centrally produced $\mathbf{K^+K^-}$ System}

  If we apply the same analysis technique to the centrally produced $K^+K^-$ system, the choice of the single physical solution becomes clearer. Only for one solution, the expected dominance of the $S$-wave at threshold is observed. In addition, this solution shows almost no intensity in the $P$-wave above the narrow $\phi$(1020), a fact that supports the assumption of double-Pomeron exchange as the dominant production process. 
For this reason, we limit the analysis to spin-0 and spin-2 contributions in the interesting mass range above $1.05\,\mathrm{GeV}/c^2$.
In this simplified case, the problem is reduced to a second-order polynomial, which leads to only two independent solutions~\cite{chu97}. Combining the information from the real and imaginary parts of the Barrelet-zeros, a unique identification along all mass bins is again possible.

\begin{figure}[t!]
  \begin{center}
    \includegraphics[width=.85\textwidth]{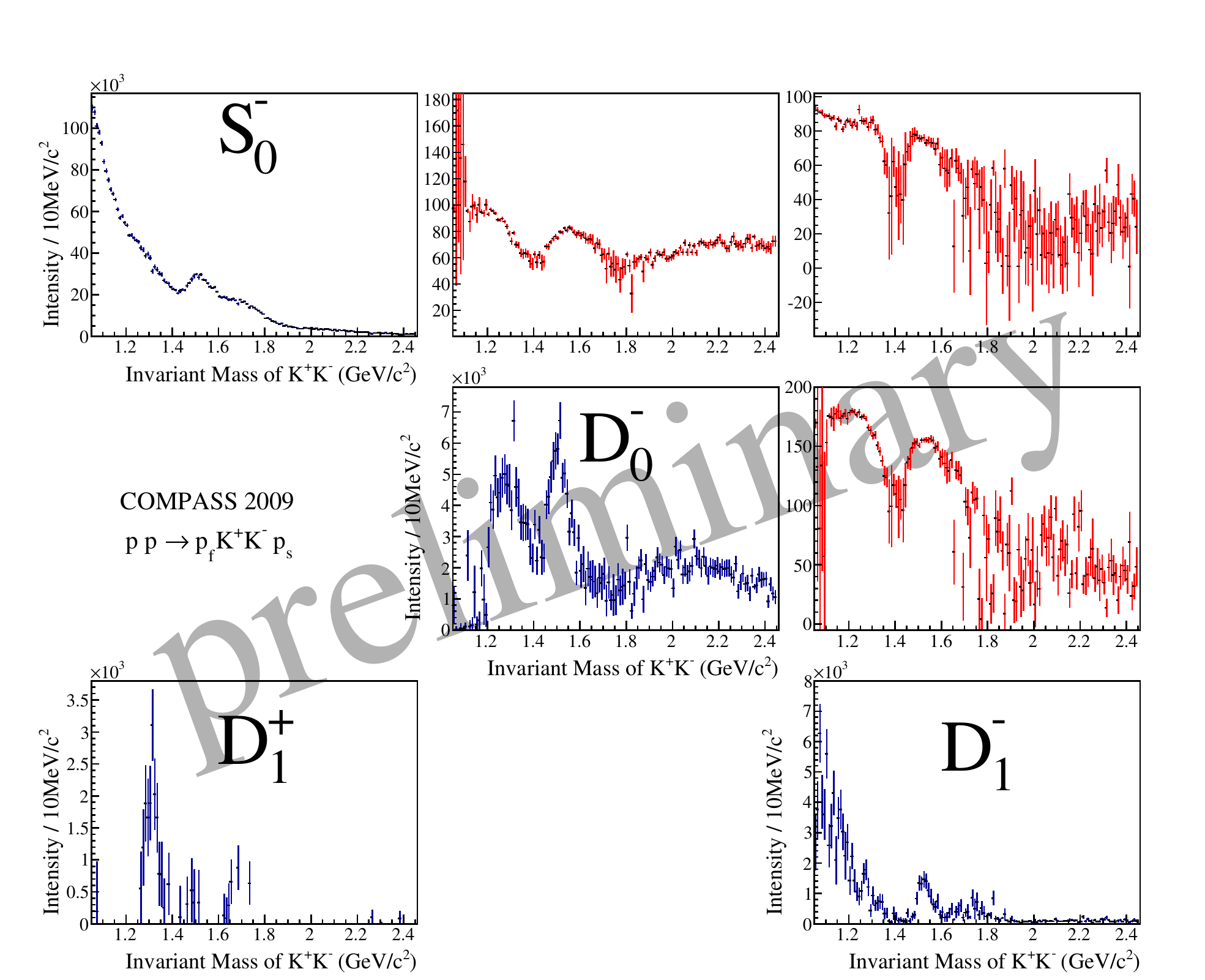}
  \end{center}
  \caption[{\em Physical solution with $S$- and $D$-waves}]{\em Physical solution with $S$- and $D$-waves, intensities in blue, phases in red.}
  \label{fig:physSD}
\end{figure}

Using the calculated physical solution as starting values, Figure~\ref{fig:physSD} is obtained by the partial-wave analysis fit in mass bins. While the broad structures in the $S^-_0$-wave are still hard to interpret at this stage, the two sharp peaks in the $D^-_0$ wave can be immediately identified as the $f_2$(1270) and the $f_2'$(1525) resonances. The latter can also be discerned in the $D^-_1$-wave with spin projection $m=1$. However, the enhancement at threshold in this wave is not expected for a spin-2 system. The distinct phase relation between $S$- and $D$-waves will give us a powerful tool to draw conclusions about the resonant contributions.


In the sector corresponding to unnatural parity-exchange, a signal for the $a_2$(1320) meson can be observed in the $D^+_1$-wave. However, the missing phase to another coherent wave does not allow to draw a conclusion here.

  \section{Mass-Dependent Parametrisation of the $\mathbf{K^+K^-}$ System}
  \label{sec:massdep}
  
  In order to be able to interpret the results of the mass-independent decomposition into partial-waves, their mass dependence has to be parametrised in terms of a physical model. The model parameters are determined by a $\chi^2$ fit to both the real and imaginary parts of the spin-density matrix elements for a subset of partial-waves. In this preliminary analysis, we focused only on the two most prominent contributions: the real (anchor) wave $S_0^-$ and the complex-valued $D_0^-$.
  
 The resonant contributions are modelled by a sum of dynamic-width relativistic Breit-Wigner functions of the following form:
 \begin{equation}
   BW(m) = \frac {\sqrt{m\Gamma(m)}} {m^2 -m_0^2 - \imath m_0 \Gamma(m)}
 \end{equation}
 where $m_0$ is the nominal mass of the resonance. $\Gamma(m)$ denotes its total width and is evaluated as
 \begin{equation}
   \Gamma(m) = \Gamma_0 \frac {q} {m} \frac {m_0} {q_0} \left ( \frac {B_{\ell}((qR)^2)} {B_{\ell}((q_0R)^2)} \right )^2
 \end{equation}
 taking into account the breakup momentum $q$ and the orbital angular momentum $\ell$ of the decay through the Blatt-Weisskopf barrier factors $B_{\ell}$~\cite{hip72} with $R=1$fm as the assumed interaction radius. $\Gamma_0$ is the nominal width of the resonance. Besides the fit parameters $m_0$ and $\Gamma_0$, we allow for a free complex amplitude for every Breit-Wigner function. 

 A coherent background had to be introduced to account for the threshold enhancement in the $S$-wave as well as for the contributions from other production processes which translate through their angular characteristics (cf.~Section~\ref{sec:weighted}) into intensity in the $D$-wave at masses above $1.8\,\mathrm{GeV}/c^2$. We parametrised the background by
 \begin{equation}
   BG(m,\ell) = (\alpha + \imath \beta) \cdot q^{\ell} \cdot \sqrt \frac{q}{m^2} \cdot \exp(-\gamma q - \delta q^2)
 \end{equation}
 with the breakup momentum $q$ and the real fit parameters $\alpha, \beta, \gamma$ and $\delta$. The factor $q^{\ell}$ provides the correct asymptotic behaviour for the angular-momentum barrier. The two-body phase space is represented by the square root term~\cite{pdg12}. Two independent background terms were introduced for the $S$- and $D$-wave. Since the overall phase is arbitrary, the background of the $S$-wave was chosen to be real ($\beta_S = 0$) in order to limit the number of fit parameters.
 
\begin{figure}[]
   \begin{center}
     \includegraphics[width=.87\textwidth]{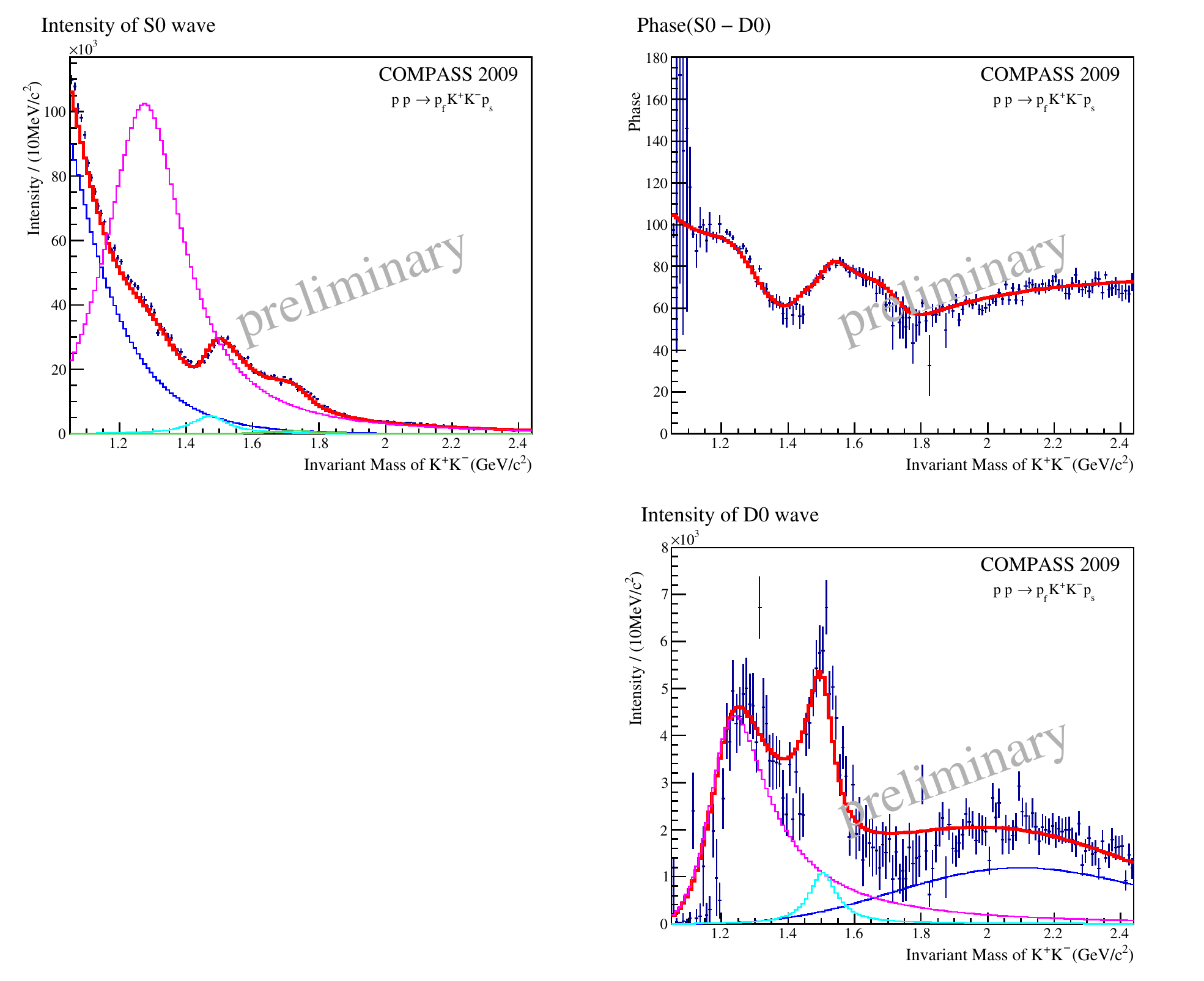}
   \end{center}
   \caption[{\em Mass-dependent parametrisation of intensities and phase}]{\em Mass-dependent parametrisation of intensities and phases (red curve) with non-resonant background (dark blue) and Breit-Wigner functions (other colours, see text)}
   \label{fig:intSD}
 \end{figure}

  
 In total, 27 parameters were used to describe the mass-dependence of the two selected partial-waves. Since the $\chi^2$-minimisation is sensitive to local minima, the fit had to be performed in several steps. At first, only the complex coupling coefficients were determined, leaving the masses and widths at the PDG~\cite{pdg12} values. In a second step, the mass parameters were released using the set of previously determined complex coefficients as the starting point. Finally, also the width parameters were determined by the fit. During this procedure, the Breit-Wigner parameters $m_0$ and $\Gamma_0$ were limited to large but non-overlapping ranges of the order of several hundred MeV$/c^2$. The final result can be seen in Figure~\ref{fig:intSD}.
 
 The two sharp peaks in the $D$-wave were fitted with two well-known resonances. One Breit-Wigner function was used to parametrise the $f_2$(1270) meson; an excited $f_2'$(1525) is used to fit the second peak in the intensity distribution. An additional $f_2$(2150) as suggested by \cite{bar99} was not needed to describe the data, the intensity can be attributed to the background. At least three different Breit-Wigner functions with different complex couplings were necessary to describe the $S$-wave. In addition to the well-established $f_0$(1500) and $f_0$(1710) resonances, a broad $f_0$(1370) had to be included to account for both the intensity as well as the phase with respect to the $D$-wave in this mass region. A strong interference with the background even required a dominant contribution from this term~(cf.~Figure~\ref{fig:intSD}). The contribution from the $f_0$(980) resonance located below the analysed mass range of $1.05 < m(K^+K^-) < 2.45\,\mathrm{GeV}/c^2$ cannot be included with a simple Breit-Wigner function. A Flatt\'e-type~\cite{fla76} parametrisation is presently studied.

The presented analysis still has a number of caveats. First of all, we restricted the mass-dependent parametrisation to $m=0$ waves so far. The inclusion of $m=1$ would add more information, but 
the unphysical rise at low masses in the intensity distribution of the \mbox{$D^-_1$-wave} indicates possible shortcomings of the simplistic model. Furthermore, the large correlation between the parameters of the $f_2$(1270) in the $S$-wave and the $f_0$(1370) in the $D$-wave can lead to systematic errors in the Breit-Wigner parameters which are under investigation at the moment. In addition, the $f_0$(1370) strength is sensitive to the background parametrisation which is under study as well. For that reason, we do not quote the mass and width parameters of the resonant contributions here.


  \section{Outlook}
  \label{sec:DaO}
  
  With these proceedings, we want to show that we are able to select centrally produced two-pseudoscalar final states and describe the main features of the data in terms of partial waves. Using the methods from~\cite{bar99}, we can reproduce the analysis with comparable results, but also with the same limitations and assumptions. 

The methods were applied to the centrally produced $\pi^+\pi^-$ and $K^+K^-$ systems. For the former, the problem of inherent mathematical ambiguities was explained, but a definite solution could not be singled out yet. The kaonic counterpart, however, allowed to simplify the model and therefore to reduce the number of mathematical ambiguities from eight to two. Subsequently, a mass-dependent parametrisation was found which could follow the intensities and phases with a minimal set of five Breit-Wigner functions and two coherent backgrounds. The obtained Breit-Wigner parameters are comparable to previous analyses and the PDG~\cite{pdg12} values. However, the systematic errors are still carefully studied. 

In the meantime, other ideas were developed to determine spin and parity of particles produced in central exclusive processes~\cite{kai03}. In addition to the measurement of the decay products, it can have particular advantages to study angular correlations between the outgoing protons. In the long run, we might study the data with a combination of both.

Nevertheless, the amount of data as well as the sensitivity of the analysis largely exceed earlier studies. The phase relations emerge with unprecedented precision and provide important information ignored by previous analyses~\cite{bar99}. The data show that COMPASS may be able to contribute to the controversial discussion about the existence of resonances in the scalar sector~\cite{och13}. In order to interpret the composition of the super-numerous scalar resonances, a combined analysis of all available final states will be essential. Especially the combination with the corresponding neutral final states $\pi^0\pi^0$, $\eta\eta$, and $K^0_SK^0_S$ can help to resolve remaining ambiguities.


\end{document}